%% file: main.tex
\documentclass[conference]{IEEEtran}
\usepackage{cite}

\ifCLASSINFOpdf
  \usepackage[pdftex]{graphicx}
\else
\fi
\usepackage{amsmath}
\usepackage{cite}
\usepackage{amsmath,amssymb,amsfonts}
\usepackage{algorithmic}
\usepackage{graphicx}
\usepackage{textcomp}
\def\BibTeX{{\rm B\kern-.05em{\sc i\kern-.025em b}\kern-.08em
    T\kern-.1667em\lower.7ex\hbox{E}\kern-.125emX}}

\usepackage[hidelinks]{hyperref}
\hypersetup{
  breaklinks=true,
}

\usepackage{eso-pic}

\usepackage{amsmath,amsfonts}
\usepackage{graphicx}
\usepackage{textcomp}
\usepackage{enumitem}
\usepackage{multirow}
\usepackage{colortbl}
\usepackage{array}
\usepackage{tcolorbox}
\usepackage{colortbl}
\usepackage{nicematrix}
\usepackage{tabularx}
\usepackage{balance}
\usepackage{booktabs}
\usepackage{url}
\usepackage{tipa}
\definecolor{lightgray}{gray}{0.9}

\PassOptionsToPackage{normalem}{ulem}
\usepackage{ulem}

\providecolor{added}{rgb}{0,0,1}
\providecolor{deleted}{rgb}{1,0,0}

\usepackage{subcaption}
\usepackage{mwe}

\usepackage{gb4e}
\noautomath

\usepackage{tikz}
\usepackage[framemethod=tikz]{mdframed}
\usetikzlibrary{calc}

\newtcolorbox{mybox}{
  sharp corners,
  colback=specialgray,
  colframe=specialblue,
  boxrule=0pt,
    toprule=0pt,
  bottomrule=0pt,
  leftrule=3pt, 
  rightrule=3pt 
}

\usepackage{booktabs}

\usepackage{cellspace}
\makeatletter
\AddToShipoutPictureBG*{%
    \AtPageLowerLeft{%
        \ifnum\thepage=1
            \put(\dimexpr(\paperwidth-\textwidth)/2\relax+8pt,35){
                \color{black}
                \makebox[\textwidth-25pt][c]{
                    \setlength{\fboxsep}{3pt}  
                    \fbox{\parbox[c]{\dimexpr\textwidth-0\fboxsep-0\fboxrule\relax}{\centering \footnotesize \textsf{© 2024 IEEE. Personal use of this material is permitted. Permission from IEEE must be obtained for all other uses, in any current or future media, including reprinting/republishing this material for advertising or promotional purposes, creating new collective works, for resale or redistribution to servers or lists, or reuse of any copyrighted component of this work in other works.}}}
                }
            }
        \fi
    }
}
\makeatother

\begin{document}
%
\title{Interlinking User Stories and GUI Prototyping: A Semi-Automatic LLM-based Approach}
%
%
%

\author{
\IEEEauthorblockN{
Kristian Kolthoff\IEEEauthorrefmark{2}\IEEEauthorrefmark{1}, 
Felix Kretzer\IEEEauthorrefmark{3}\IEEEauthorrefmark{1}, 
Christian Bartelt\IEEEauthorrefmark{2}, 
Alexander Maedche\IEEEauthorrefmark{3},
and Simone Paolo Ponzetto\IEEEauthorrefmark{4}}

\IEEEauthorblockA{\IEEEauthorrefmark{1}
Authors contributed equally to the paper.}

\IEEEauthorblockA{
\IEEEauthorrefmark{2}Institute for Enterprise Systems,
University of Mannheim, Mannheim, Germany\\
Email: \{kolthoff, bartelt\} @es.uni-mannheim.de}

\IEEEauthorblockA{\IEEEauthorrefmark{3}
Human-Centered Systems Lab,
Karlsruhe Institute of Technology, Karlsruhe, Germany\\
Email: \{felix.kretzer, alexander.maedche\} @kit.edu}

\IEEEauthorblockA{\IEEEauthorrefmark{4}
Data and Web Science Group, 
University of Mannheim, Mannheim, Germany\\
Email: simone@informatik.uni-mannheim.de}}

%
%

\markboth{2024 IEEE 32nd International Requirements Engineering Conference (RE)}%
{Interlinking User Stories and GUI Prototyping: A Semi-Automatic LLM-based Approach}
%



\maketitle

\begin{abstract}
\input{Chapters/00_Abstract}
\end{abstract}


%
\IEEEpeerreviewmaketitle

\input{Chapters/01_Introduction}

\input{Chapters/02_RelatedWork}
\input{Chapters/03_Methodology}
\input{Chapters/04a_Evaluation}
\input{Chapters/04b_Results}
\input{Chapters/05_Discussion}
\input{Chapters/06_Conclusion}

\ifCLASSOPTIONcaptionsoff
  \newpage
\fi



%
\newpage
\bibliographystyle{IEEEtran}
\bibliography{bibtex/bib/IEEE_Bib_File}

\end{document}

%% file: Chapters/00_Abstract.tex
Interactive systems are omnipresent today and the need to create graphical user interfaces (GUIs) is just as ubiquitous. For the elicitation and validation of requirements, GUI prototyping is a well-known and effective technique, typically employed after gathering initial user requirements represented in natural language (NL) (e.g., in the form of \textit{user stories}). Unfortunately, GUI prototyping often requires extensive resources, resulting in a costly and time-consuming process. Despite various easy-to-use prototyping tools in practice, there is often a lack of adequate resources for developing GUI prototypes based on given user requirements. In this work, we present a novel Large Language Model (LLM)-based approach providing assistance for validating the implementation of functional NL-based requirements in a GUI prototype embedded in a prototyping tool. In particular, our approach aims to detect functional user stories that are not implemented in a GUI prototype and provides recommendations for suitable GUI components directly implementing the requirements. We collected requirements for existing GUIs in the form of user stories and evaluated our proposed validation and recommendation approach with this dataset. The obtained results are promising for user story validation and we demonstrate feasibility for the GUI component recommendations.

%% file: Chapters/01_Introduction.tex
\section{Introduction}

Graphical user interfaces (GUIs) have become ubiquitous, allowing users to interact with software applications in most aspects of our daily lives. This trend has led to an increasing demand for GUIs. Today, providing GUIs that meet requirements is an essential commercial success factor for software applications \cite{jansen_graphical_1998}. In practice, agile methods have increasingly been used instead of traditional phase-based methods \cite{caseStudyAgile2011}. \textit{Agile requirements engineering} attempts to address the changes that agile methods bring with them, such as requirements engineering and design activities being carried out continuously throughout development projects \cite{caseStudyAgile2011, caseStudyAgile2015}. This can also lead to a deeper integration of stakeholders into the development process, promising better overall results (\hspace{1sp}\cite{BRHEL2015163, McKeen_1994, Vredenburg2002, mao_2005}).

A popular technique to involve stakeholders in the development phases and facilitate reflection on requirements, is \textit{prototyping} of GUIs. The use of prototypes in requirements elicitation was already investigated more than two decades ago \cite{Mannio2001RequirementsEU}. Recent analyses \cite{slrEilicitationTechniques} have shown that prototypes serve for "efficient feedback and collaboration among stakeholders" \cite[p. 372]{slrEilicitationTechniques}, as a tool to reflect on collected requirements and as a catalyst reducing elicitation time \cite{slrEilicitationTechniques}.

However, GUI prototyping often requires substantial resources \cite{shams2009requirements} making it a costly and time-consuming process (e.g., because creating GUI prototypes can require knowledge in interface design and programming \cite{RuddStern_lowHighFidelity}). 
A particular challenge, while using GUI prototypes for requirements elicitation, results from the iterative change of requirements (e.g., user stories) in requirements elicitation. Debnath et al. \cite{Sourav_RE_USCHANGE}, as an example, present a study, where less than half of the later user stories "include content that can be fully traced to the initial ones" \cite[p. 233]{Sourav_RE_USCHANGE}, and a high percentage of resulting user stories were new or refinements of the initial ones. Due to the iterative change of formalized requirements, GUI prototypes are often redesigned with new or changed requirements. Time and effort lie in recognizing whether requirements have already been implemented and then implementing new requirements in GUI prototypes.

While others have looked into supporting users tasked with creating GUI prototypes from different perspectives, to the best of our knowledge, there exists no approach that automatically checks requirements (e.g., in the popular form of user stories) against implemented components in GUI prototypes and provides recommendations for not-implemented requirements that can directly be integrated. With recommendations for improvements, users tasked with creating GUI prototypes can be enabled to create more effective GUI prototypes for requirements elicitation and validation. While different NL-based GUI retrieval strategies (e.g.,\cite{bernal2019guigle, kolthoff_data-driven_2023, kolthoff2020gui2wire, kolthoff2021automated}) were proposed in this area, those GUI retrieval strategies mainly aim to generate first GUI inspirations and cannot be utilized to automatically assess whether or to what degree a GUI prototype meets single requirements. Work on generating images of GUIs from textual descriptions (e.g., using stable diffusion \cite{UI_Diffuser_2023}) often comes with the limitations inherent to its output format – images – namely challenges in assessing the implementation of individual user stories on images, and complicated processing of images in follow-up prototyping steps, since images cannot be modified with  prototyping tools from practice such as Figma \cite{figma}. Assistants tailored at supporting prototyping within dedicated tools (e.g., GUIComp \cite{guiComp2020}) either provide examples based on initial GUI prototypes as source of inspiration or optimize prototypes with design metrics. Those tools are not primarily connected to context-dependent requirements and, to some degree, do not consider how users can effectively translate requirements into a (initial) prototype.

In order to address the presented research gap, we explore the question of \textit{how to effectively detect functional user stories not implemented in GUI prototypes and provide recommendations for suitable GUI components}? With our approach, we investigate how prototypes can be effectively aligned with functional user stories, e.g., to be used later for reflection with stakeholders (e.g., as throwaway or evolutionary prototypes). For our approach, we focus on directly linking requirements and prototypes and consider an initial set of requirements (formalized as user stories) as given. We decided for user stories in our approach since they represent a popular formalization of requirements in practice, and literature \cite{slrEilicitationTechniques}. 

We contribute by \textit{first} proposing and evaluating an approach detecting functional user stories not implemented in GUI prototypes and providing recommendations for suitable GUI components directly implementing the requirement, \textit{second} by outlining a system implementing our approach and by providing a research plan on how to evaluate the proposed system, and \textit{third} by making our code, dataset, and material needed to reproduce our approach and foster future research publicly available at \cite{paper_repository}.

%% file: Chapters/02_RelatedWork.tex
\section{Related Work}

Various approaches in previous research address automated support for requirements engineering, validation, or GUI prototyping in general (for an overview, see, e.g., \cite{METH20131695, umar_advances_2024}).
Umar and Lano \cite{umar_advances_2024} present a summary of automated support for requirements engineering. They note that most automated tools for requirements elicitation support aim to create \textit{Unified Modelling Language} (UML) from less structured requirements. The automated creation of UML differs significantly from our approach since UML is less intuitive and more complex for stakeholders. Furthermore, UML lacks the capabilities to visualize basic functionality and interactions in direct comparison to GUI prototypes. Therefore, UML can present a challenge when eliciting and validating requirements with stakeholders, whereas GUI prototypes are suitable.

Prior research such as \textit{Guigle} \cite{bernal2019guigle}, \textit{GUI2WiRe} \cite{kolthoff2020gui2wire, kolthoff2021automated} and \textit{RaWi} \cite{kolthoff_data-driven_2023} presented NL-based GUI retrieval strategies exploiting the large-scale GUI repository \textit{Rico}. Their approaches primarily focus on the exploration and assessment of diverse techniques for NL-based GUI retrieval, with the aim for providing GUI design ideas or useful support for requirements analysts in a requirements elicitation context, respectively. In contrast, our proposed approach supports requirements analysts while creating GUI prototypes on a finer-grained user story level. Moreover, our recommendation approach provides support for custom requirements compared to the restriction to the available repository of retrieval-based approaches. 
Furthermore, numerous GUI retrieval methods leveraging visual input have been previously suggested. Swire \cite{huang2019swire}, for instance, exploits visual embeddings as a means to retrieve GUIs from hand-drawn sketches. GUIFetch \cite{behrang2018guifetch}, on the other hand, offers retrieval of comprehensive applications based on exhaustive Android application sketches. 
Moreover, VINS \cite{bunian2021vins} advocates GUI retrieval employing either a rudimentary wireframe prototype or a fully implemented GUI prototype as input. GUIComp \cite{guiComp2020} polls similar GUIs from a finite set of pre-build GUIs based on initial GUI prototypes. While these approaches can support requirements analysts during the GUI prototyping phase, they merely support sketches as input and therefore neglect NL requirements in the form of user stories that often are gathered in the initial requirements elicitation phase. 

Additionally, with the arrival of generative AI in many research areas, tools like UI-Diffuser \cite{UI_Diffuser_2023} allow the fast generation of GUI prototypes based on prompts using stable diffusion. However, approaches like UI-Diffuser come with limitations inherent to the output format: images. Generated images can currently not serve as a basis to automatically detect whether all requirements have been implemented in the generated images. In addition, there is a disconnection from prototyping in practice, as images cannot be used as input for prototyping tools (e.g., Figma \cite{figma}). Therefore, even minor adjustments cannot be made in familiar prototyping tools.

Some approaches have already investigated automated testing of user stories on GUI prototypes. Silva et al. \cite{silva_10.1007/978-3-030-24289-3_46} present a Behaviour-Driven Development (BDD) based approach that enables the automated testing of user stories on interactive GUI prototypes with web browser automation tools. However, the interactive GUI prototypes required for this are often already further developed and not in the scope of rapid prototyping to elicit and validate requirements. Furthermore, in said approach no recommendations are generated or directly implemented into GUI prototypes in comparison to our proposed approach.

%% file: Chapters/03_Methodology.tex
\section{Approach}

This section provides an overview of our proposed approach to support prototype developers during the creation of GUI prototypes from user stories. Requirements elicitation typically starts with an elicitation interview with stakeholders and initial NL requirements are often gathered in the form of user stories and cleansed afterwards \cite{schneider2007generating}. Subsequently, initial low-fidelity GUI prototypes are created often using GUI prototyping tools, e.g. Figma \cite{figma}. Depending on the problem, GUI prototypes are created from scratch or based on templates (e.g. retrieved by \textit{RaWi} \cite{kolthoff_data-driven_2023}) that already fractionally match the gathered user stories. In such a scenario, our approach aims to support the prototype developer by \textit{(i)} validating the current GUI prototype state against the user story collection to show missing user stories and \textit{(ii)} providing implementation recommendations in the form of visualized GUI-DSL (Domain-Specific Language) for the user stories. Our approach is divided into several main components as shown in Fig. \ref{fig:overview}. First, \textit{(A)} a GUI prototype abstraction component to transform the DSL of the GUI prototyping editor to an abstracted textual representation, \textit{(B)} an user story validation component that utilizes the GUI abstraction and user story collection in an LLM-based approach to classify weather a user story is already implemented, \textit{(C)} a component matching the GUI components to implemented user stories and \textit{(D)} a recommendation component to provide GUI suggestions of how the user story could be implemented.

\subsection{GUI Prototype Abstraction}

As an input to the previously mentioned LLM-based methods, the GUI prototype requires to be transformed to a simplified abstract textual representation. Typically, prototyping tools employ a custom DSL or object model to hierarchically represent the GUI prototype. To enable the different prediction and recommendation tasks, we focus on extracting merely functional aspects from the prototypes, i.e. component types, their displayed texts, their names providing important semantic information and their boundaries. Reducing the extracted information to the mentioned aspects helps in both reducing consumed context length in the LLMs and focusing the model solely on functional aspects. Currently, the approach is not fully integrated into a GUI prototyping tool. Therefore, in our preliminary evaluation we employ the \textit{Rico}\cite{RICO_Deka_2017} GUI repository to obtain initial results, as we can similarly extract all the mentioned aspects from the semi-automatically gathered GUIs. For each extracted GUI component, we then create a textual representation using the following abstract pattern followed by three examples created from \textit{Rico} GUIs:

\begin{center}
\textit{"uicomp-text"} \textbf{(uicomp-type)} \texttt{(uicomp-name)} \\
\textit{"+7.10"} \textbf{(Label)} \texttt{(price Change TV)}\\
\textit{"Install App"} \textbf{(Button)} \texttt{(native Ad Call To)} \\
\textit{"Example: 'New York'"} \textbf{(Text Input)} \texttt{(location)}
\end{center}

\begin{figure}[t]
  \includegraphics[width=0.47\textwidth]{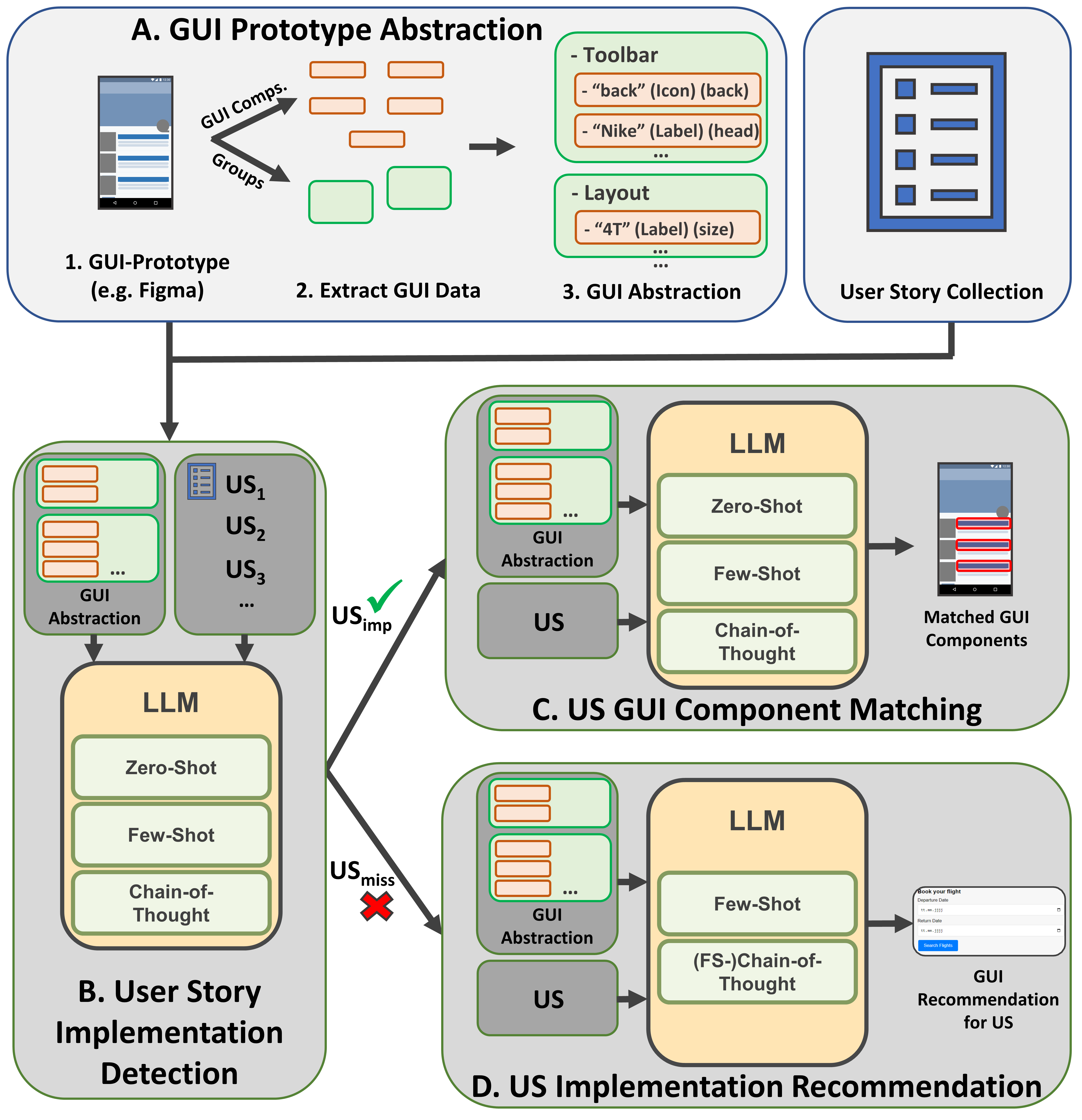}
  \caption{Overview of our proposed approach pipeline (A-D)}
	\label{fig:overview}
\end{figure}

\noindent Specifically, the \textit{uicomp-text} refers to the displayed text of the component, the \textbf{uicomp-type} refers to the basic GUI component type (e.g. Label, Button, Checkbox etc.) and the \texttt{uicomp-name} refers to the name of the component given within the prototyping editor. In the absence of any of the components, the respective field is left empty. This textual representation of the GUI components encompasses relevant information from a functional perspective. Moreover, we derive clustering elements from the semantic annotations of \textit{Rico}, which incorporate categories such as \textit{List Item}, \textit{Card}, and \textit{Toolbar}, among others. These represent names of layout groups and can similarly be extracted from a GUI prototyping editor. To identify clusters for the remaining components not encompassed by the preceding groups, we further extracted layout clusters from the original GUI hierarchy by aligning them with the GUI components. Subsequently, we fabricate the GUI representation as two-tier bullet points, with the outer tier representing the layout groups and the inner tier denoting their corresponding GUI components. Prior to generating the string representation, the layout groups are arranged based on their boundaries from the top-left to the bottom-right, and in a similar fashion, the GUI components within each group are organized shown in Fig. \ref{fig:overview} to resemble the original GUI layout.

\subsection{User Story Implementation Detection}

To tackle the problem of identifying whether a user story is implemented in a GUI prototype, we propose several LLM-based methods and approach it as a binary classification problem. The recent surge in popularity of LLMs can be attributed to their capacity for swift learning and adaptation to novel tasks, relying solely on a limited number of examples \cite{brown2020language, kojima2022large}. These models are particularly versatile, capable of being tailored to a wide array of specific tasks through a method known as \textit{In-Context Learning (ICL)} or \textit{prompting} \cite{liu2023pre}, respectively. Given the extensive knowledge encapsulated within LLMs and the access to this knowledge via prompting, the integration of these models for the detection and recommendation problems at hand are promising.

In particular, we adopt the \textbf{Zero-Shot (ZS)} prompting method \cite{kojima2022large} by creating a prompting template divided into \textit{(i)} a task instruction providing clear guidelines for the model, \textit{(ii)} the user story to validate followed by \textit{(iii)} the generated GUI abstraction. We instruct the model to predict a single token for the classification and we extract the log probabilities for both labels, which provides a user story ranking mechanism. In particular, the extracted probability can be employed to estimate the certainty of the classification. This probability can be further exploited to rank the user stories from high to low probabilities (e.g., for later visualization to users).

In addition, we adopt the \textbf{Few-Shot (FS)} prompting method \cite{brown2020language} showing often enhanced performance for various tasks. In \textbf{FS} prompting, we basically follow the \textbf{ZS} pattern, however, we additionally provide several \textit{input-output} pairs to guide the model for the specific task. For our preliminary evaluation, we created multiple \textbf{FS} prompting templates (varying examples).

Moreover, we adopt the \textbf{Chain-of-Thought (COT)} prompting method \cite{wei2022chain}, in which the LLM is instructed to create multiple intermediate reasoning steps before generating a prediction. Thus, we instruct the model to first generate an explanation providing reasoning whether the user story is implemented. In addition, this provides an interpretable explanation that can further be used for later error analysis.

\subsection{User Story GUI Component Matching}

In addition to solely predict the coverage of a user story in a GUI prototype, we further investigate the task of extracting all GUI components from the prototype that are required to fulfill the user story. This represents a natural extension of the previous task, enabling direct interlinking the user story with its respective GUI components and gaining deeper insights in the LLMs predictions. Similarly to the previous task, we adopt \textbf{ZS}, \textbf{FS} and \textbf{CoT} prompting models for this matching task. However, we extend the GUI abstraction by adding identifiers to each GUI component and correspondingly adapt the instructions and prompting templates to enable the LLM to output a parsable collection of GUI component identifiers.

\subsection{User Story Implementation Recommendation}

Finally, in order to not only interlink user stories with the GUI prototype and detect missing user stories but also directly support the prototype developer to implement missing user stories in the GUI prototype, the last component in our pipeline represents a LLM-based recommendation approach. In particular, we adopt \textbf{FS} and \textbf{FS-CoT} prompting models to generate a ranking of possible implementations of the user story contextualized based on the current GUI prototype state. In our preliminary investigation of these methods, we decided to generate recommendations in the form of HTML/CSS due to \textit{(i)} the employed LLMs typically being pretrained on large amounts of HTML documents (generate error-free syntax) and \textit{(ii)} the easy visualization of the generated recommendations. For the integration of these methods into a GUI prototyping tool, we aim for generating a DSL syntactically close the editor-DSL facilitating the integration of the recommendation into the GUI. This can be achieved either by fine-tuning the LLM on the DSL or potentially solely by employing \textit{ICL}, as shown earlier for a DSL in robotic planning environment \cite{wake2023chatgpt}.

\subsection{Proposed Integrated System}

\begin{figure}
    \includegraphics[width=8.87cm]{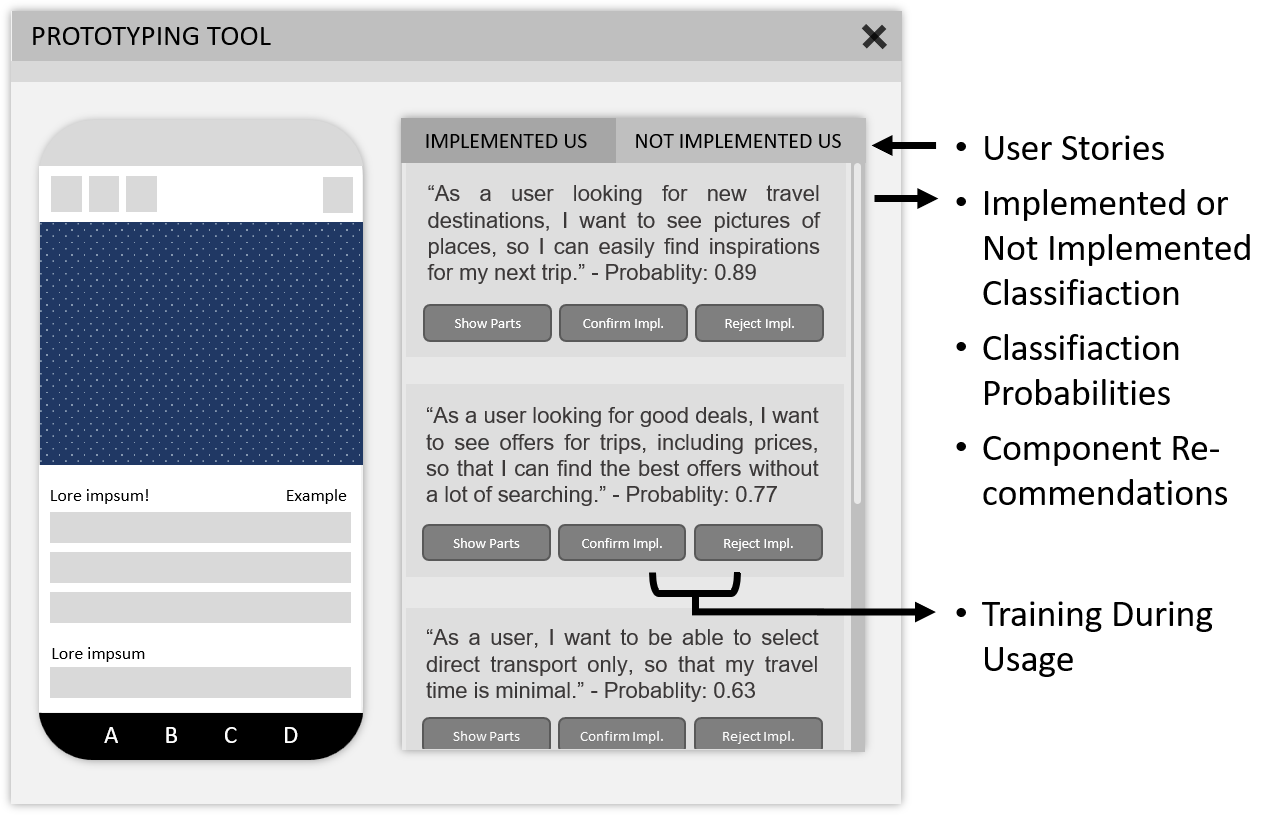}
    \centering
    \caption{Proposed system integrating user story based validation and GUI recommendations into dedicated prototyping tools}
    \label{fig:integratedSystem}
\end{figure}

In this section, we propose an integrated system (illustrated in Fig. \ref{fig:integratedSystem}) to demonstrate the implementation of our approach. In line with \cite{Kretzer10.1145/3544549.3585659}, we propose integrating our approach directly into dedicated prototyping tools (such as, e.g., Figma \cite{figma} or Adobe xD \cite{adobe_inc_adobe_2023}). The direct integration - e.g., in the form of a plug-in - supports the rapid creation of prototypes directly in the appropriate tools and, simultaneously, makes it possible to access the DSL required for our approach (illustrated in Fig. \ref{fig:overview}). We propose four main features: First, \textit{(i)} a list of all user stories classified as \textit{implemented} and \textit{not-implemented} may be displayed directly next to the GUI prototype. In addition, communicating the probability with which the respective user story was classified may support users by communicating uncertainty.
\textit{Second}, \textit{(ii)} with direct integration, users can let our system highlight all GUI components matching a single user story classified as implemented and thereby gain an understanding of the classification and which components are related to a particular feature.
\textit{Third}, \textit{(iii)} users can contribute to continuous learning and fine-tuning our approach with integrated user feedback for each user story, e.g., by marking incorrectly classified user stories as such in our system. 
\textit{Fourth}, \textit{(iv)} our system's recommendations - generated in the respective DSL - can be directly integrated into a GUI prototype since our system can interact with the prototypes in dedicated prototyping tools. In this regard, our system can potentially reduce resource consumption when creating initial prototypes.

%% file: Chapters/04a_Evaluation.tex
\section{Evaluation}

This section delineates the design of our preliminary evaluation of the proposed approach. Since the approach is not yet fully implemented, we focus the evaluation on two main aspects of the approach including the user story implementation detection and GUI component matching methods. For enabling this preliminary evaluation, we constructed a gold standard of user stories and GUI prototype annotations. To this end, we formulate the subsequent two research questions:

\begin{itemize}[leftmargin=0in]
    \item[] \textbf{RQ$_{1}$}: How effective are LLM-based approaches for detecting user story implementation in GUI prototypes?
    \item[] \textbf{RQ$_{2}$}: How effective are LLM-based approaches for extracting GUI components fulfilling a user story from a GUI prototype?
\end{itemize}

\subsection{Data Collection}
To evaluate our approach, a dataset of GUIs and associated user stories was required. While there are established datasets of GUI prototypes available (e.g. \textit{Rico} \cite{RICO_Deka_2017}), to our knowledge there exists no dataset combining GUI prototypes with user stories. We therefore decided to collect user stories for existing GUI prototypes from \textit{Rico}. The following section introduces how we collected and preprocessed the dataset by presenting existing GUIs to study participants creating the user stories.

\textit{GUI Sample}. In order to get a broad selection of different \textit{Rico} GUIs, our initial GUI sample was randomly drawn from ten different domains. Following our exclusion criteria, we then selected valid GUIs from our random sample. We decided ex-ante to exclude interfaces with non-English text, personal data displayed, overlays (such as pop-ups) shown, components without annotations in the \textit{Rico} dataset, trivial GUIs (e.g., simple log-in screens resulting in the same repetitive user stories), and to exclude interfaces with unclear functionality. Our final sample included 60 GUIs from the domains: \textit{Shopping} (8), \textit{Health \& Fitness} (11), \textit{Education} (5), \textit{News} (4), \textit{Sports} (6), \textit{Travel} (6), \textit{Books} (5), \textit{Music} (6), \textit{Finance} (4), and \textit{Food \& Drink} (5). We additionally created GUI versions where each component was annotated with a number so that participants could assign their user stories to one or more GUI components.

\textit{Procedure and Survey}. User stories were collected using a questionnaire. The participants were presented with information about the study, data protection, and the conditions of participation. They then learned how to write functional user stories. Learning content was supported with examples of user stories and concluded with comprehension checks. Participants had two chances for each comprehension check before a screenout took place. Participants then created user stories for nine consecutively shown GUIs. We instructed the participants to create three to five functional user stories describing features already implemented and provided a user stories template for guidance. After creating user stories for nine GUIs, the participants' final task was to specify for each user story which components of the GUI belong to the respective user story. The participants were shown one after the other the identical nine GUIs with the already created user stories, but the GUIs now contained numbers that annotated each GUI component.

\textit{Participants}. We selected 8 participants (2 female, 6 male) from a student pool. Participants were on average 24.3 years old, had 1.1 years of experience creating and 1.0 years experience in evaluating visual designs, such as GUI prototypes. Each participant generated on average 4.5 user stories per GUI.

\textit{Data Processing}. Overall, the participants created 327 user stories, with duplicate user stories and varying quality (e.g., despite explicit instructions, some user stories were written for \textit{not-implemented} features). In order to obtain a usable dataset, the user stories were cleaned up. Therefore, two paper authors labeled each user story independently. For this purpose, the first step was to determine whether the user story \textit{a)} fully meets the requirements, \textit{b)} contains one or more errors, or \textit{c)} is a duplicate of a previous user story. After the separate labeling, the inter-coder reliability was calculated (Cohen's Kappa $\kappa = .548$). After resolving disputes, 231 user stories (with their respective GUI prototypes) were included in the final data set, whereas 96 user stories (16 duplicates, 80 different exclusion criteria) were not considered further. The applied exclusion criteria are provided in our accompanying repository \cite{paper_repository}.

\subsection{RQ$_{1}$: User Story Implementation Detection}

To answer RQ$_{1}$, we evaluated the ability of various LLM-based prompting techniques to predict whether a user story is contained in a GUI prototype. To this end, we created a gold standard based on the collected user stories and GUI annotation pairs. First, we randomly selected 5 GUIs comprising 21 user stories to be employed as examples for the \textbf{FS} prompting. The remaining 210 US-GUI-pairs form the basis for the gold standard. This procedure ensures the avoidance of overlapping GUI abstraction data between gold standard and \textbf{FS}-examples and introducing bias. Next, we randomly assigned half of the examples (105) the class \textit{Implemented (1)} and the remainder the class \textit{Not-Implemented (0)}. While the GUI data for the first class remains unchanged, in the GUI abstraction of the second class the paired GUI component annotations were removed. For example, consider the second GUI of Fig. \ref{fig:matching_examples}. For the shown US, the respective GUI components associated with the US according to the gold standard are marked in the GUI. To include such an US as a negative example in the gold standard, we would remove the respective GUI components from the GUI abstraction (two labels and a checkbox). We compute precision ($P$), recall ($R$), F1-measure ($F1$) and accuracy ($ACC$). To conduct the experiments, we employed the most recent \textit{GPT-4} model\cite{openai2023gpt4} (8,192 tokens context length, \textit{temperature=0}, accessed in February 2024) as our base LLM, holding the benchmark on many NLP tasks. For the \textbf{FS} prompting, we evaluated one model with five (\textbf{FS$_{5}$}) and another with ten examples (\textbf{FS$_{10}$}), respectively. In addition, we evaluated four \textbf{CoT} models with varying \textit{temperature}. This is based on the idea that with varying \textit{temperature}, we restrict or allow the model to provide more or less diverse explanations.

\subsection{RQ$_{2}$: User Story GUI Component Matching}

To answer RQ$_{2}$, we evaluated the ability of several LLM-based prompting techniques to extract all GUI components relevant to fulfill a given user story. Therefore, we employed the same gold standard as previously described, however, using the original unchanged abstraction for each of the 210 GUIs and the labels being the annotated GUI component identifiers from the gold standard. Precisely, as the input, the model received the GUI abstraction (each GUI component marked by an ID) and a US to predict a set of GUI components IDs relevant for the US. Next, we compared the set of extracted GUI component identifiers with the gold standard set of identifiers and computed $P$, $R$ and $F1$ measure. We computed these metrics for each example in the dataset and averaged them over the gold standard to obtain \textit{macro} values. Therefore, \textit{macro} values represent the average of each metric over the gold standard. Although metric values for two US examples might be equal, the absolute amount of correct or erroneous classifications might differ significantly between US depending on the GUI component set length. For example, the second US example from Fig. \ref{fig:matching_examples} is only associated with three respective GUI components, whereas another US from the gold standard about visualizing an overview of the daily nutrients consumed daily (see gold standard example 78 \cite{paper_repository}) has 20 associated GUI components. To take into account these set length differences across the gold standard and thus counter potential inaccurateness introduced by set length, we additionally constructed binary prediction arrays to compute respective \textit{micro} values possessing correct weights i.e. double GUI component set length leads to double the influence on the final metric result. To conduct the experiments, we employed the identical setup of LLMs as described previously for RQ$_{1}$.

%% file: Chapters/04b_Results.tex
\section{Results and Discussion}

In this section, we briefly present the preliminary evaluation results and provide answers to our guiding research questions.

\subsection{RQ$_{1}$: User Story Implementation Detection}

Table \ref{tab:results_rq1} illustrates the evaluation results for RQ$_{1}$, showing the $P$, $R$ and $F1$ metrics for both classes and the $ACC$ over the created gold standard. First, we can observe a substantially high absolute performance across all of the investigated prompting methods indicated by, for example, $ACC$ scores of .852 (\textbf{ZS}), .848 (\textbf{FS$_{5}$}) and .829 (\textbf{CoT$_{t=1}$}). This indicates that LLMs are capable of effectively processing the semantics of the created GUI abstraction and match it to the semantics of the functionality encompassed in the user stories. These high metric values indicate that LLMs can produce promising results for the approach. Although the \textbf{ZS} method seems to perform best overall, the respective pairwise \textit{McNemar} tests between each of the seemingly best performing models of each prompting method indicates no statistical significance. Moreover, the \textbf{CoT} methods apparently tend to be more restrictive about predicting that a user story is implemented, as indicated by the highest $P_{1}$ and $R_{0}$ values. In contrast, the \textbf{ZS} and \textbf{FS} methods appear to be more balanced among the metrics and classes.

\begin{table}[!t]
\footnotesize
\caption[Evaluation Results]{Evaluation results of various LLM-based prompting approaches for US validation in GUI prototypes (Binary class.)}
\centering
\setlength\tabcolsep{6pt} 
\begin{tabular}{Sl|ccc|ccc|c}
\toprule
& \multicolumn{3}{c|}{\textbf{US Implemented (1)}} & \multicolumn{3}{c|}{\textbf{Not-Implemented (0)}} & \multicolumn{1}{c}{\textbf{}}\\
 \cmidrule(lr){2-4} \cmidrule(lr){5-7} \cmidrule(l){8-8} 
& $P_{1}$ & $R_{1}$ & $F1_{1}$ & $P_{0}$ & $R_{0}$ & $F1_{0}$  & $ACC$   \\ \midrule
\textbf{Zero-Shot}      &.830           &.886       &.857   &.878          &.819       &.847   &.852 \\ \midrule
\rowcolor{lightgray}\textbf{Few-Shot$_{5}$}     &.829          &.876       &.852   &.869          &.819       &.843   &.848  \\
\textbf{Few-Shot$_{10}$}      &.818          &.857       &.837   &.850           &.810        &.829   &.833  \\ \midrule
\rowcolor{lightgray}\textbf{CoT$_{t=0}$}      &.900            &.686       &.778   &.746          &.924       &.826   &.805  \\
\textbf{CoT$_{t=.5}$}      &.888          &.676       &.768   &.738          &.914       &.817   &.795  \\ 
\rowcolor{lightgray}\textbf{CoT$_{t=1}$}      &.888          &.752       &.814   &.785          &.905       &.841   &.829  \\
\textbf{CoT$_{t=1.3}$}     &.812          &.743       &.776   &.763          &.829       &.795   &.786 \\ \midrule
\bottomrule
\end{tabular}
\label{tab:results_rq1}
\end{table}

To enhance the understanding of the misclassifications made by the LLM, we conducted an error analysis and investigated the $FP$ and $FN$ instances. For the $FP$ instances, the main root cause for misclassifications appears to be a semantic misinterpretation of GUI components with reference to the user story by the LLM. For example, for a user story that describes to provide addresses of the nearest stores the LLM identifies the component \textit{"SAN FRANCISCO, Store \#6498"} \textbf{(Button)} \texttt{(storelocator address line1)} as fulfilling the user story, although this component merely provides the city name and the detailed address fields were absent. Similarly, for the user story to enable/disable the ability to mark days as complete within a settings GUI of a fitness app, the LLM identified the GUI component \textit{"check"} \textbf{(Icon)} \texttt{(done)} as fulfilling the user story. However, this component is located in the GUI toolbar and refers to saving the overall settings.

For the $FN$ instances, we identified several similar main root causes. Often, detailed semantic information about the functionality of the GUI components might be absent in the created GUI abstraction resulting in the LLM being restrictive about positively identifying the user story as fulfilled. For example, a user story requiring a donation button to easily donate money could not be detected, since the GUI component was implemented as an image without any further textual description, hence, the image information is not accessible to the model. Similarly, a user story to add new shopping lists to a collection of lists could not be identified since the description of the GUI component \textit{"add"} \textbf{(Icon)} \texttt{(overview viewpager fab)} is general and ambiguous for detection.

\subsection{RQ$_{2}$: User Story GUI Component Matching}

\begin{table}
\footnotesize
\caption[Evaluation Results]{Evaluation results of various LLM-based prompting approaches for (comp.) matching of US in GUI prototypes}
\centering
\setlength\tabcolsep{6pt}
\begin{tabular}{Sl|ccc|cccc}
\toprule
& \multicolumn{3}{c|}{\textbf{Macro}} & \multicolumn{4}{c}{\textbf{Micro}} \\
 \cmidrule(lr){2-4} \cmidrule(l){5-8}
& $P$ & $R$ & $F1$ & $P$ & $R$ & $F1$  & $ACC$   \\ \midrule
\textbf{Zero-Shot$_{A}$}      &.784            &.819         &.755     &.620             &.755         &.681     &.516 \\
\rowcolor{lightgray}\textbf{Zero-Shot$_{B}$}     &.718            &.903         &.743     &.502            &.858         &.633     &.463  \\ \midrule
\textbf{Few-Shot$_{5}$}     &.850             &.755         &.765     &.677            &.643         &.659     &.492 \\ \midrule
\rowcolor{lightgray}\textbf{CoT$_{t=0}$}      &.758            &.806         &.727     &.557            &.731         &.632     &.462  \\
\textbf{CoT$_{t=.5}$}      &.721            &.788         &.688     &.492            &.710          &.581     &.409  \\ 
\rowcolor{lightgray}\textbf{CoT$_{t=1}$}      &.698            &.809         &.690      &.534            &.746         &.622     &.452  \\
\textbf{CoT$_{t=1.3}$}    &.677            &.740          &.654     &.562            &.646         &.601     &.430 \\ \midrule
\bottomrule
\end{tabular}
\label{tab:results_rq2}
\end{table}

\begin{figure}[t]
      \centering
      \includegraphics[width=0.485\textwidth]{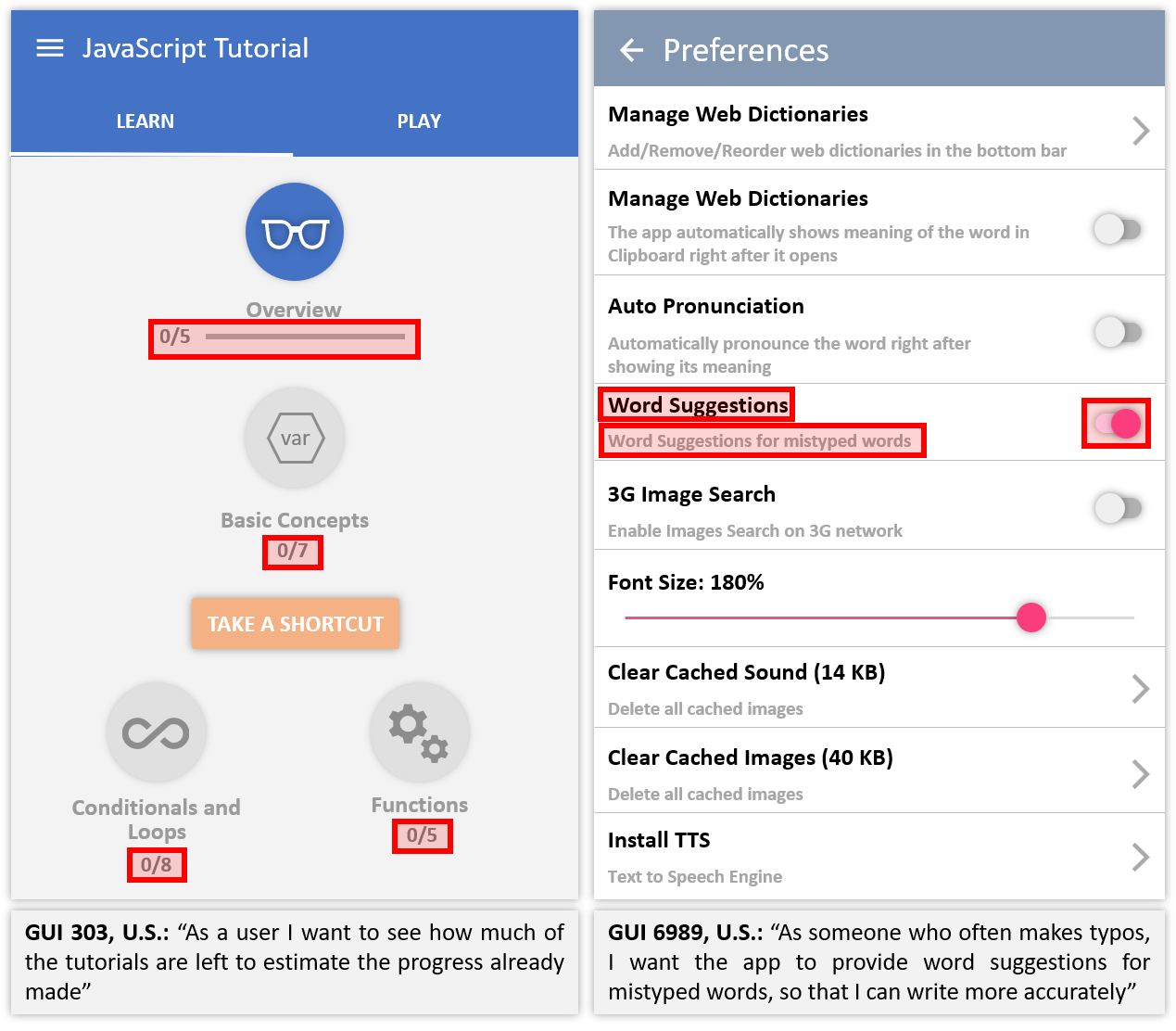}
      \caption{GUI component matches to user stories. Redrawn from Rico \cite{RICO_Deka_2017} using components from \textit{Material Design} \cite{google_llc_material_2024}}
    \label{fig:matching_examples}
\end{figure}

Table \ref{tab:results_rq2} illustrates the evaluation results for RQ$_{2}$, showing the macro and micro $P$, $R$ and $F1$ metrics and the micro $ACC$ values. Overall, the results indicate that the models achieve a moderate to good performance of matching GUI components to user stories shown by, for example, \textit{Micro}-$F1$ scores of .681 (\textbf{ZS$_{A}$}), .659 (\textbf{FS$_{5}$}) and .632 (\textbf{CoT$_{t=0}$}). Although a direct comparison with the results of the task discussed in RQ$_{1}$ is difficult due to the difference in datasets, still the matching task can be seen as an extension of the classification task, since classification models probably perform similar computations (e.g., as indicated by explanations from \textbf{CoT} models) as part of their reasoning sequence. As can be observed, the performance difference of the tasks indicates that the matching task is significantly more difficult for the LLMs. However, we argue that the obtained results are promising due to the model being capable of extracting the majority of GUI components correctly shown by the metric values. As indicated by the Wilcoxon-signed-ranked test between the \textit{Macro}-$F1$ scores of the method, the \textbf{CoT} prompting methods perform significantly worse, whereas the differences of the \textbf{ZS} and \textbf{FS} methods are insignificant. In addition, the \textbf{ZS$_{B}$} prompting method has significantly better $R$ values compared to \textbf{ZS$_{A}$} (in vice versa for $P$ values), indicating that the prompt extension in \textbf{ZS$_{B}$} optimizes the model for not missing component matches. Fig. \ref{fig:matching_examples} shows example GUIs from the gold standard and highlighted GUI component matches as generated by the LLM.

\begin{figure}
    \centering
    \includegraphics[width=0.475\textwidth]{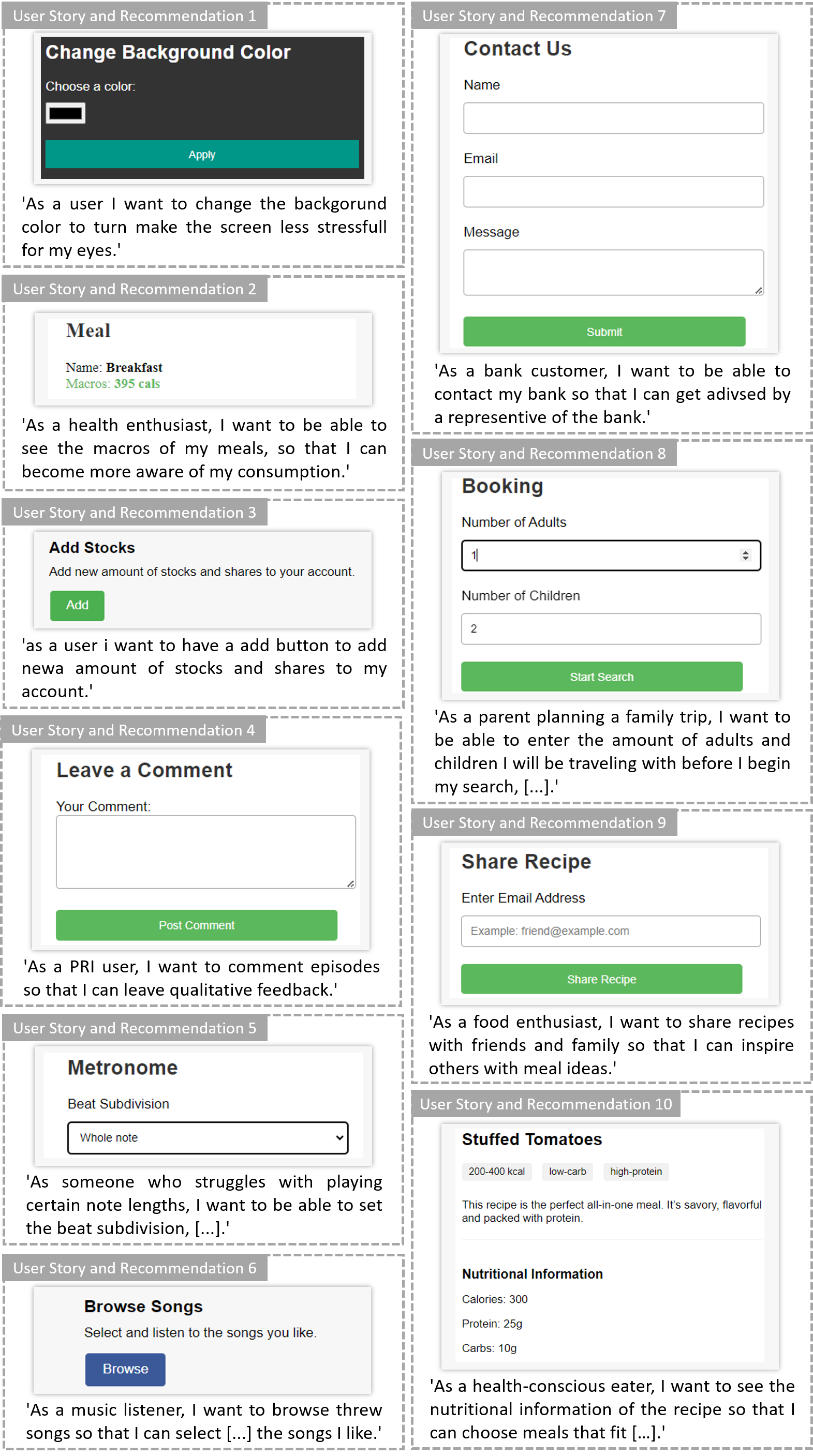}
    \caption{GUI feature recommendations (randomly sampled)}
    \label{fig:recommendation_examples}
\end{figure}

Moreover, we investigated \textit{low} $P$ and/or \textit{low} $R$ instances to improve the understanding of errors made by the LLM. For the cases of \textit{low} $P$, the LLM often extracted wrong GUI components that were semantically related. For example, for the user story to specify the number of guests in a hotel search GUI, the models also erroneously extracts components for the number of rooms. For the instances possessing \textit{low} $R$, similar to the misclassifications discussed earlier, we identified missing or ambiguous descriptions of GUI components as a main cause. For example, for a user story to be able to download videos to watch them offline, the GUI components offering download functionality were represented as \textbf{(Image)} \texttt{(document)}. Although the succeeding GUI component \textit{"0.7 MB"} \textbf{(Label)} \texttt{(document size)} could act as a hint, the naming as a \textit{document} and the component being marked as an image introduces ambiguity. Finally, some cases possess ambiguity that is difficult to resolve without the stakeholder. For example, for the user story requesting to see an overview of the course (Fig. \ref{fig:matching_examples}) the LLM extracts all listed lessons, whereas the annotation marks only the overview course item.

\subsection{Preliminary Recommendation Results}
\label{Recommendations}

Due to the early stage of the proposed approach, we did not yet fully evaluate the recommendation performance of LLM-based prompting techniques. However, we provide preliminary results for the recommendation task in the following. To this end, we generated recommendations (HTML/CSS) with both \textbf{FS} and \textbf{FS-CoT} prompting models for all user stories in our gold standard. Fig. \ref{fig:recommendation_examples} shows ten randomly sampled recommendations generated by the \textbf{FS} model and their respective US. For example, for the first US the model creates all required GUI components i.e. a \textit{color picker}, an \textit{apply button} and the respective \textit{labels}. In addition, consider the fifth example, for which the model not only recommends a reasonable main GUI component (\textit{drop down selection}), but also pre-fills it with matching domain information, exploiting the wide domain and general knowledge embedded in LLMs. Due to the small amount of provided few-shot examples, the designs of the generated recommendations appear repetitive. However, since our focus primarily lies on generating the functionality required for the US, this represents only a minor issue. In the future, this could be mitigated by providing more few-shot examples and fine-tuning. We generated the recommendations for both \textbf{FS} and \textbf{FS-CoT} methods and all user stories of our gold standard and provide them as HTML documents and image visualizations in our repository \cite{paper_repository}.

%% file: Chapters/05_Discussion.tex
\section{Limitations and Potential Risks} 

Reflecting on our approach and the proposed integrated system, we identified several limitations and potential risks for our research plan. In the following, we highlight some key limitations, potential risks, and mitigation strategies.

\textit{Evaluation dataset}. \textit{Rico} is widely adopted in research and thus indicates suitability for GUI-based evaluation tasks. Nevertheless, the dataset is influenced by the collection methods. Rico GUIs present a sample of free mobile applications from the Google Play store. Therefore, our used data set lacks non-mobile GUIs and has a tendency for GUIs based on Google's material design system \cite{google_llc_material_2024}. Additionally, to our knowledge there exists no dataset combining GUI prototypes with user stories. We therefore decided to collect user stories for existing GUIs, which does not represent the natural order of requirements elicitation. The exhaustiveness of user stories per GUI in our data set may be limited, as a fixed number of user stories were written by 8 participants for the respective GUIs. Despite independent labeling, calculating inter-coder reliability and resolving conflicts, a further limitation might be introduced by the authors' evaluation of collected user stories. We also focused on functional user stories first. To expand our dataset, we plan to collect user stories from participants before the corresponding GUIs are developed.

\textit{Prototyping tool integration.}
In our proposed integrated system, our approach is directly integrated into dedicated prototyping tools. This requires a vica-versa translation between the DSL of the prototyping tool and our textual representation. We have examined translations for Figma \cite{figma} and discuss the resulting limitations in the following paragraph. Overall, our approach can be integrated into different prototyping tools, but also works directly with markup language for web browsers (i.e., HTML and CSS, where a translation to our textual representation is feasible). Direct integration into prototyping tools may limit the operational capabilities of our approach. However, tool integration offers advantages such as the fast and cost-effective creation of GUI prototype iterations to communicate with stakeholders before more resource-intensive development steps (e.g. in HTML and CSS) follow.

\textit{Data quality in prototyping tools}. While we consider user stories and GUI prototypes for our approach as given, our approach depends on the quality of both. Prototyping tools from practice do not strictly enforce the GUI qualities (e.g., suitable naming of the components, or that parts of a single component are grouped together). Often prototyping tools use a tree structure for ordering components and allow that parts, e.g., of a button (lines, labels, or colored areas), being arranged in confusing parts of the underlying tree. Low quality inputs affect recognition and recommendations.
We aim at mitigating the risk of fluctuating input quality in our system. Besides educating participants when creating GUI prototypes, we propose pre-built components (e.g., a button that has pre-grouped and pre-labeled parts). Pre-built components are also standard in practice since corporate design often finds its way into prototyping tools through pre-built GUI design libraries.

\textit{Evaluating recommendations}. While we have yet to evaluate the recommendations generated by our approach, we have generated a broad sample of user story-based recommendations. Some of the recommendations are also described in this paper in section \ref{Recommendations}. However, an in-detail evaluation, as part of our research plan, will be addressed in future work.

\textit{Generating initial GUI prototypes}.
Our approach aims at detecting whether user stories are present in GUI prototypes. Consequently, creating initial GUI prototypes is a natural step before matching user stories. We have yet to evaluate how effective our proposed system can generate recommendations before initial GUI prototypes are build. However, we are confident that our approach can create effective recommendations for a blank canvas based solely on user stories.

%% file: Chapters/06_Conclusion.tex
\vspace{0.05cm}
\section{Research Plan and Conclusion}
In this paper, we present an approach and integrated system for detecting functional user story implementation in GUI prototypes and providing GUI component recommendations. To further evaluate our approach and the integrated system, we plan multiple next research steps:

\textit{Evaluating recommendations}. Our preliminary evaluation demonstrated the feasibility of LLMs to generate relevant GUI component recommendations for a given user story. However, we have yet to evaluate these recommendations on a broad scale. As one of the following steps, we plan to evaluate the recommendations regarding perceived usability by prototyping experts, along with the proposed research question of how effective LLM-based recommendations from functional user stories are for improving prototypes. In particular, we plan to conduct a user-based evaluation by collecting relevance annotations for LLM-generated GUI feature implementation recommendations for a larger set of user stories.

\textit{Expanding our dataset}. As described in the limitations, we asked participants to create user stories for existing GUIs. While the feasibility of our approach can already be demonstrated based on this data, we plan to collect data in a natural sequence next. The initial elicitation will precede the creation of initial GUI prototypes.

\vspace{0.01cm}
\textit{Evaluating proposed integrated system}. 
Our proposed system integrates our approach into dedicated prototyping tools, enabling a holistic evaluation of our approach. As part of our research plan, we plan to evaluate the entire system in a controlled between-subjects user study while asking how effective an integrated prototyping system is in detecting user story implementation and providing recommendations.

\vspace{0.02cm}
Although there are still unanswered research questions for our approach and integrated system, our preliminary results not only show promising potential for the user story validation and GUI component matching tasks, but also demonstrate feasibility for the GUI feature recommendation task. In addition, the future of our approach holds further potential in supporting requirements elicitation, validation, and overall improvements for users prototyping GUIs.